\documentclass[showpacs,twocolumn]{revtex4}
\usepackage{graphicx}
\usepackage{amsmath}

\begin{document}
\title{
Cumulative distribution functions associated with bubble-nucleation processes in cavitation
}

\author{Hiroshi Watanabe\footnote{watanabe@cc.u-tokyo.ac.jp}$^1$, Masaru Suzuki$^2$, and Nobuyasu Ito$^3$}

\affiliation{
$^1$Supercomputing Division,
Information Technology Center,
The University of Tokyo, 2-11-16 Yayoi, Bunkyo, Tokyo 113-8658, Japan
}

\affiliation{
$^2$ Department of Applied Quantum Physics and Nuclear Engineering, Kyushu
University, 744 Motooka, Nishi-ku, Fukuoka 819-0395, Japan
}

\affiliation{
$^3$ Department of Applied Physics, School of Engineering,
The University of Tokyo, Hongo, Bunkyo-ku, Tokyo 113-8656, Japan
}

\begin{abstract}
Bubble-nucleation processes of a Lennard-Jones liquid are studied by molecular dynamics simulations.
Waiting time, which is the lifetime of a superheated liquid, is
determined for several system sizes, and the apparent finite-size effect of the nucleation rate is observed.
From the cumulative distribution function of the nucleation events,
the bubble-nucleation process is found to be not a simple Poisson process but
a Poisson process with an additional relaxation time.
The parameters of the exponential distribution associated with the process are determined by
taking the relaxation time into account, and the apparent finite-size effect is removed.
These results imply that the use of 
the arithmetic mean of the waiting time until a bubble grows to the critical size
leads to an incorrect estimation of the nucleation rate.
\end{abstract}


\maketitle

\section{Introduction}
When a system exhibits a first-order transition
at some set of parameter values, nucleation phenomena
are observed in a transient process. 
A familiar example of such nucleation is observed in the liquid-gas phase transition.
The theory describing such nucleation was initiated by Gibbs~\cite{Gibbs1878},
which was followed by quantitative arguments for the steady-state nucleation rate~\cite{Volmer1926, Farkas1927}.
This nucleation theory was refined by Zeldovich~\cite{Zoldovich1942}
and is now called the classical nucleation theory (CNT)~\cite{Feder1966}.
CNT was first constructed for droplet nuclei in a supersaturated vapor
and then applied to bubble nuclei in a superheated liquid~\cite{Blander1975}.
In CNT, the nucleation is treated as a stochastic process, and CNT
predicts a nucleation rate that corresponds to the emerging frequency of critical embryos.
While nucleation rates during the homogeneous condensation of
a supersaturated vapor are, on the whole, well predicted by CNT with some modifications~\cite{Horsch2008},
it is well known that the nucleation rates of bubbles in a superheated liquid predicted by CNT are markedly different from the values
obtained from experimentally~\cite{Vinogradov2008} or numerically~\cite{Yamamoto2010}. 
This discrepancy originates from the fact that the nucleation phenomena of bubbles are
considerably different from those of droplets for the following reasons.
(i) The free energy required to form a bubble is not a unique function of its volume or the number of particles since a gas is compressible.
(ii) Work carried out by bubbles on the ambient liquid cannot be ignored.
(iii) Interbubble interactions via ambient liquid cannot be ignored. One of the important differences
between the droplet and bubble nucleation is the density of the ambient phase.
For the case of the droplet-nucleation, the ambient phase is gas phase which density is usually negligible.
However, the bubbles can interact each other via ambient liquid. For example,
one bubble growth increases the pressure of the surrounding liquid which may
suppress the nuclei of other embryos.
Corresponding to (i), the compressibility of bubbles can be measured by molecular dynamics (MD) simulations~\cite{Kinjo1998}, and
the free energy surface of the embryos of bubbles has recently been studied
in terms of density functional theory (DFT)~\cite{Uline2007}.
The work carried out by bubbles can also be estimated directly by MD~\cite{Yamamoto2010} and
indirectly by DFT from the analysis of cavity~\cite{Punnathanam2003}.
Despite such studies, less attention has been paid to the finite-size effect on bubble-nucleation,
while such effect was investigated for droplet-nucleation~\cite{Chkonia2009}.
If the effect from (iii) is exhibited, this would cause strong finite-size effect.
Therefore, investigation of the size dependence of nucleation rates is necessary.
In the present article, we study bubble-nucleation processes of a Lennard-Jones liquid by MD to clarify
the finite-size effect of bubble nuclei.

\begin{figure}[tb]
\begin{center}
\includegraphics[width=\linewidth]{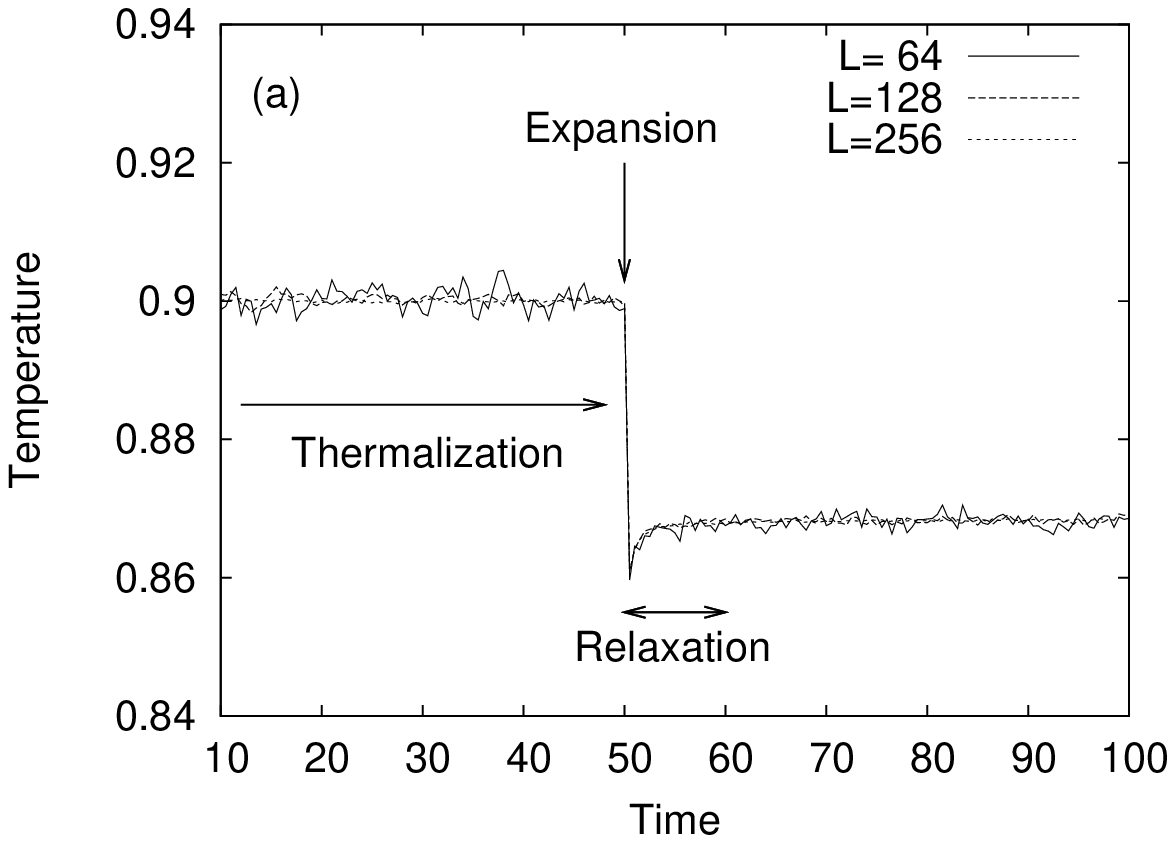}
\includegraphics[width=\linewidth]{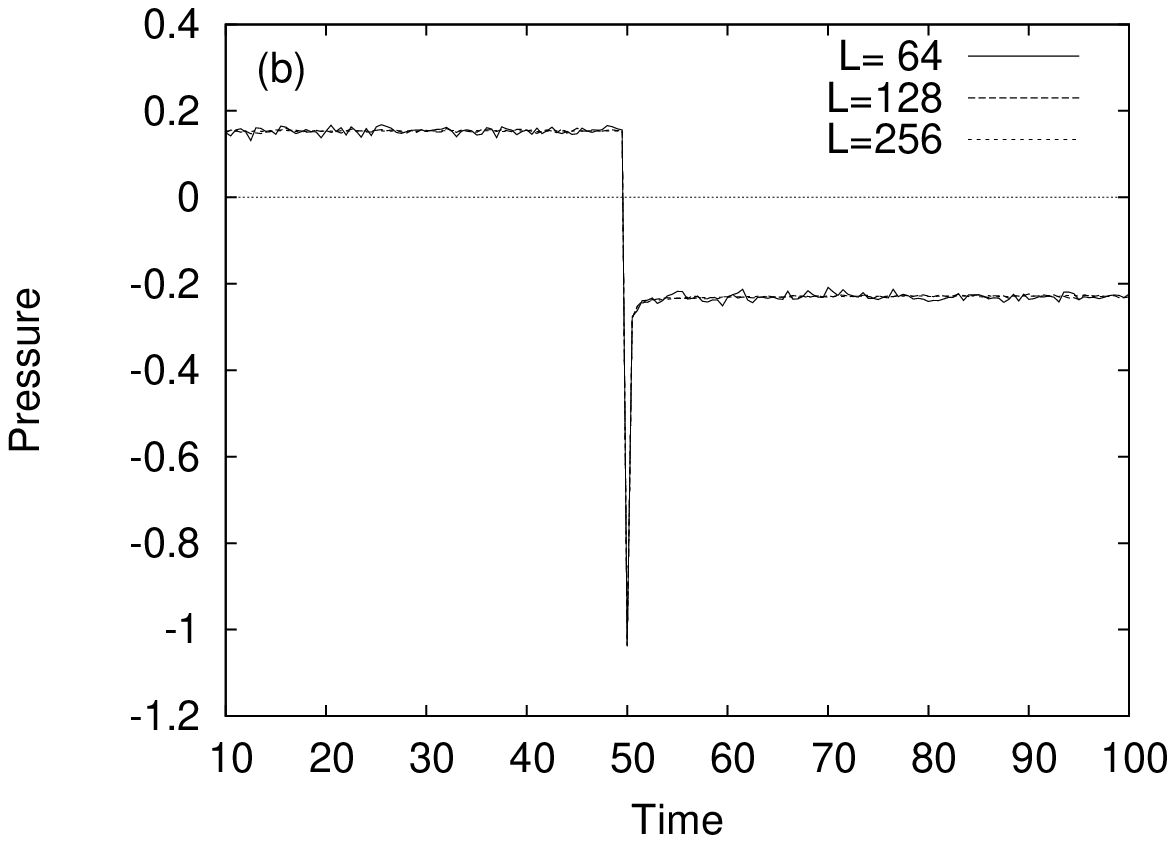}
\end{center}
\caption{
Time evolutions of (a) temperature and (b) pressure.
Only the data for $L=64, 128,$ and $256$ are shown for visibility.
After thermalization at $T=0.9$, density suddenly decreases from $0.7$
to $0.659$ at $t=50$. Then the system relaxes to a metastable state with $T=0.868(2)$.
The time evolutions of different system sizes are almost identical.
}
\label{fig_temperature}
\end{figure}

\begin{figure}[tb]
\begin{center}
\includegraphics[width=\linewidth]{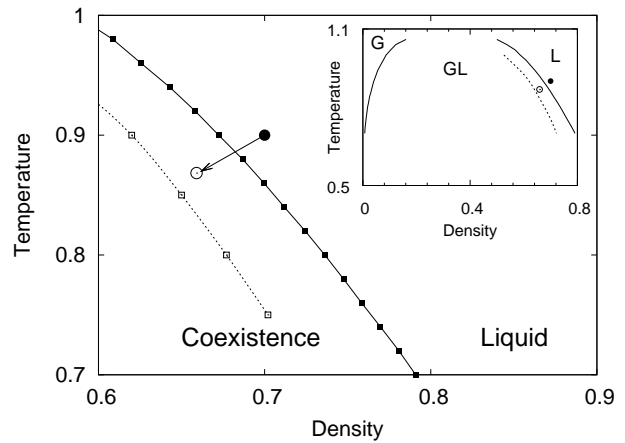}
\end{center}
\caption{
Phase diagram showing the vicinity of the phase boundary between the pure liquid and gas-liquid coexistent regions.
The solid line denotes the binodal line between the liquid and gas-liquid coexistent phases, and
the dashed line denotes the spinodal line where spinodal decomposition takes place.
The filled and open squares on the binodal and spinodal lines denote the observed point
which are used to draw the lines.
The system is first thermalized in the pure-liquid phase, then is suddenly expanded.
Then the system crosses the binodal line and becomes a metastable state (superheated liquid)
in the vicinity of the spinodal line.
The solid and open circles denote the states of the system before and after expansion, respectively.
(Inset) Overall view of the gas-liquid phase diagram. The symbols `G', `L', and `GL' denote
the pure gas, pure liquid, and gas-liquid coexistent regions, respectively. 
}
\label{fig_phasediagram}
\end{figure}

\begin{figure*}[tb]
\begin{center}
\includegraphics[width=0.32\linewidth]{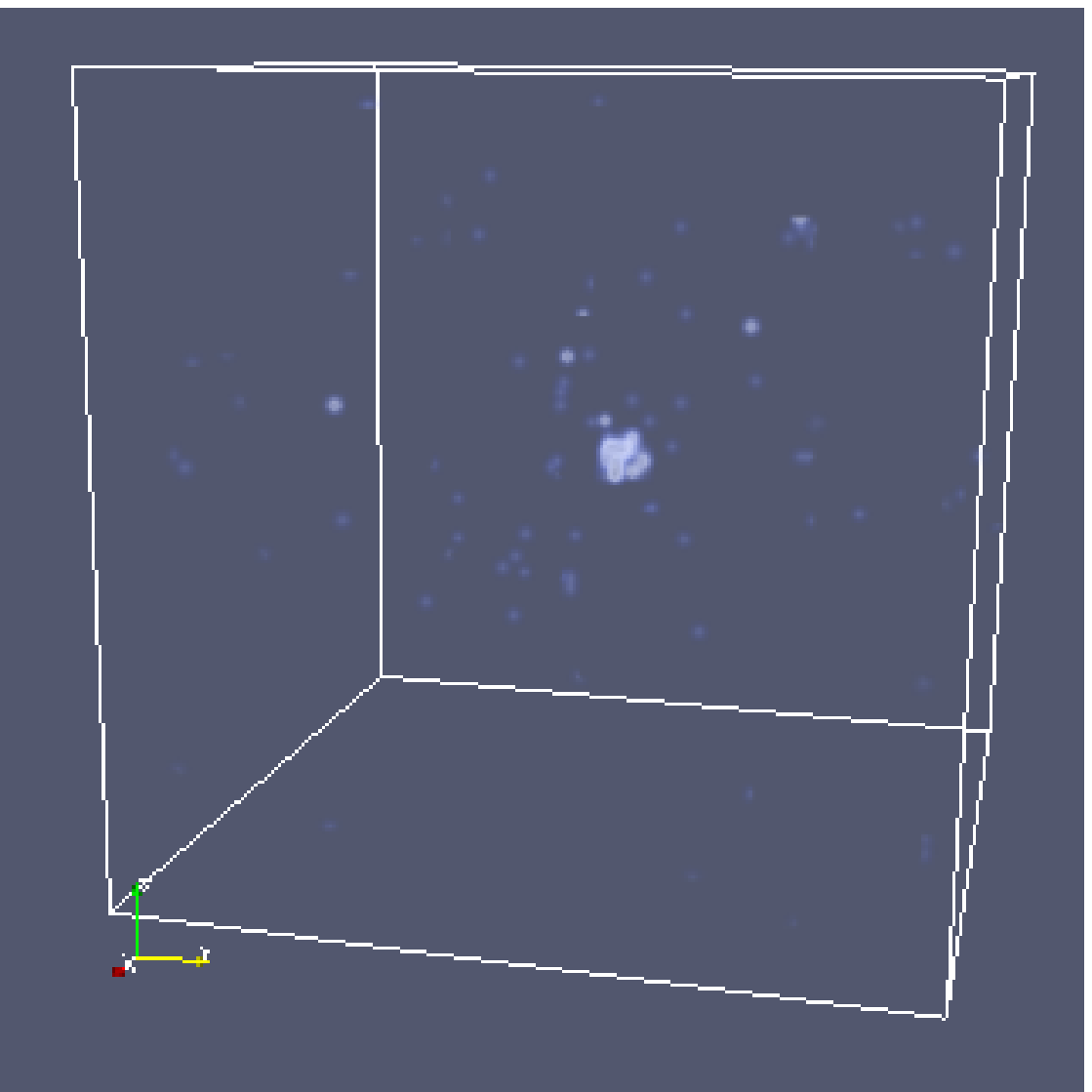}
\includegraphics[width=0.32\linewidth]{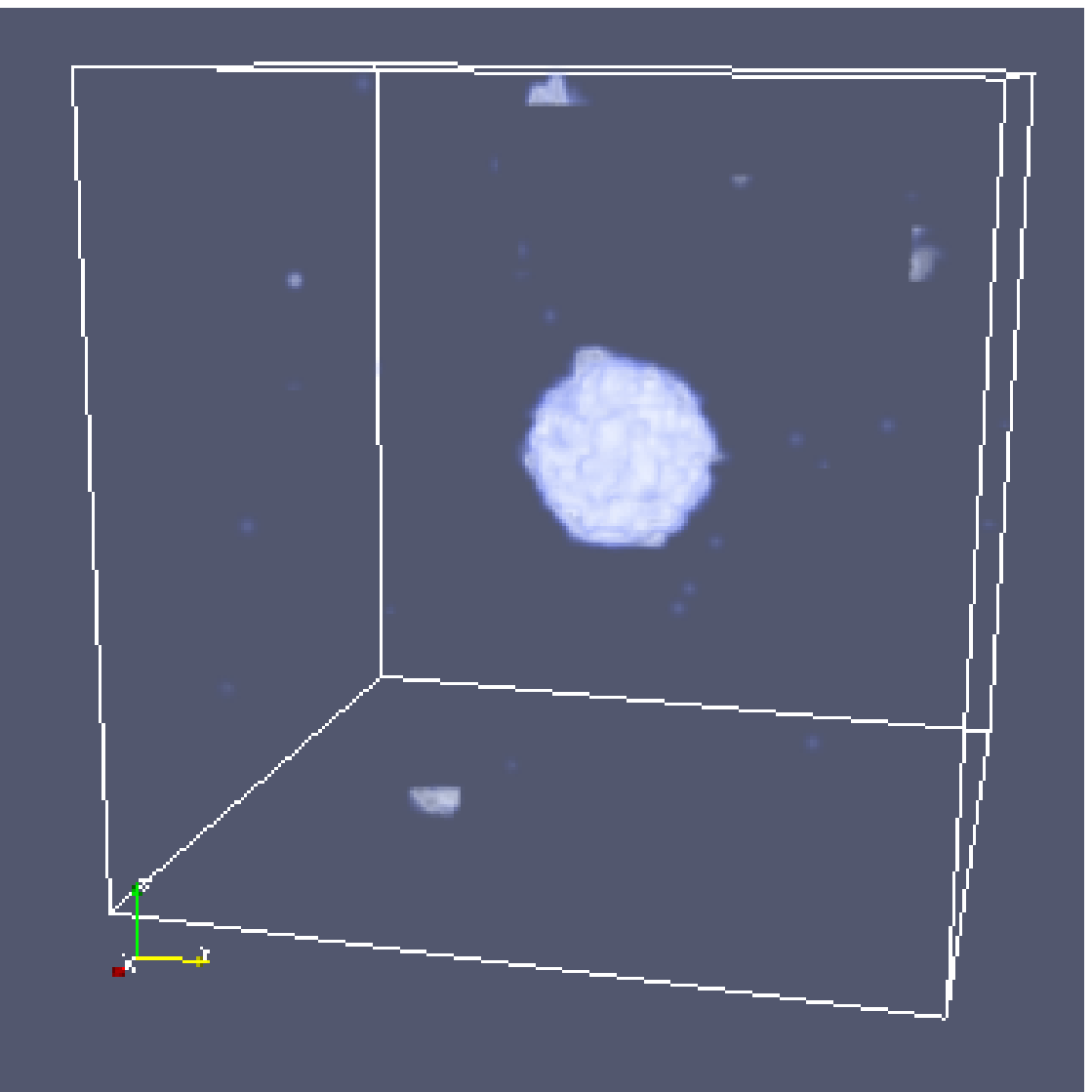}
\includegraphics[width=0.32\linewidth]{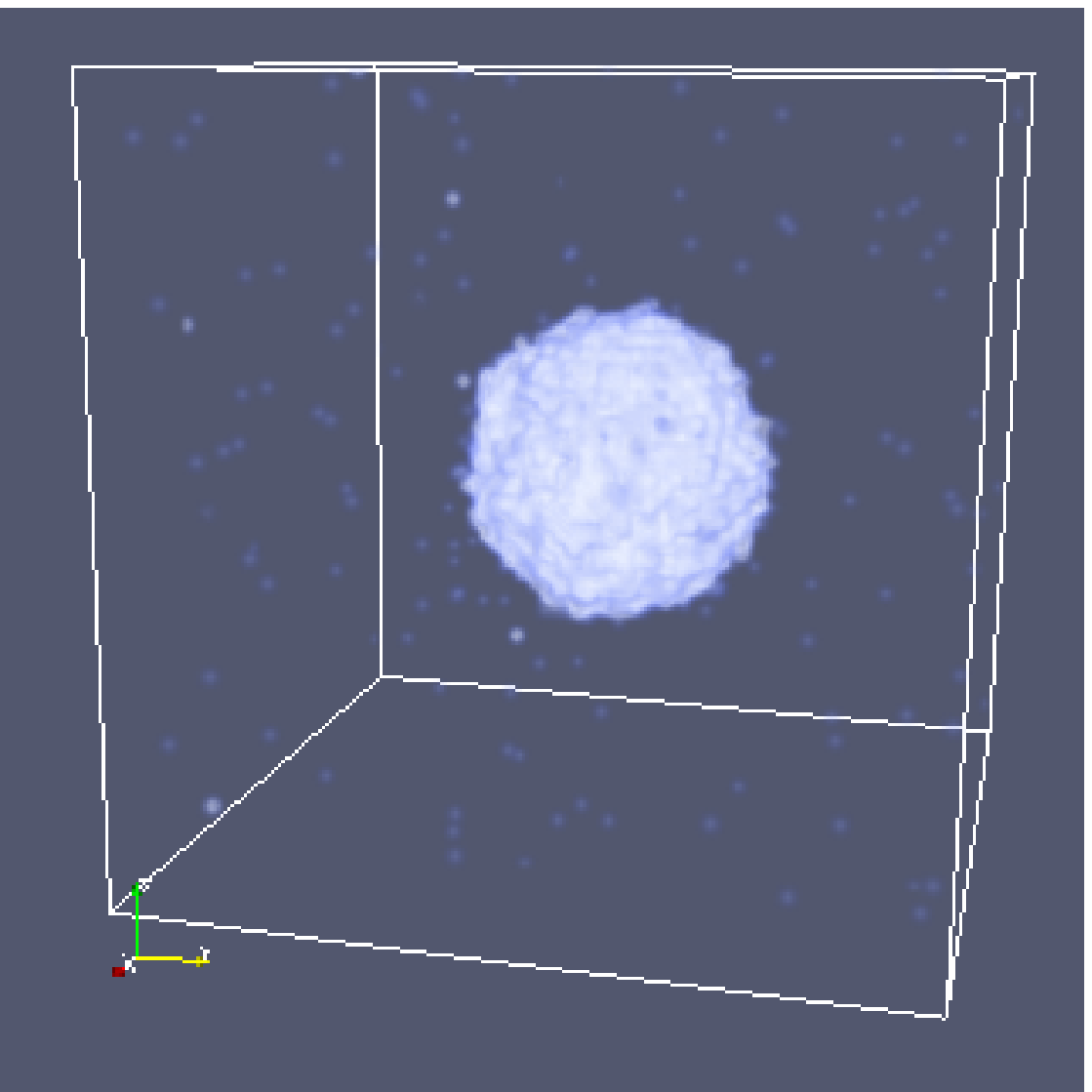}
\end{center}
\caption{
(Color online) Time evolution of bubble in the system with $L=128$.
Left to right: snapshots at $t=100$, $150$, and $550$.
We divide the system into small subcells and define them to be in the \textit{gas state} when their densities are less than $0.2$.
Only the \textit{gas state} subcells are shown.
An embryo appears after a waiting time then grows slowly into a large bubble.
}
\label{fig_nucleation}
\end{figure*}

\section{Nucleation Rate}

Consider a superheated liquid in a metastable state.
While the gas phase is more stable than the liquid phase,
time is required for the uniform phase to change to the gas-liquid coexistent phase
since there is an energy barrier to be overcome in the formation of large bubbles.
The lifetime of such a superheated liquid, which we call the waiting time $t_\mathrm{w}$ in the following,
is a stochastic variable.
The superheated liquid is characterized by a nucleation rate $J$,
which is the number of embryos growing beyond the critical size in the unit time and volume.
If the system is in the steady state, $J$ can be expressed in terms of
the expectation value of the waiting time $\left< t_\mathrm{w} \right>$ as
\begin{equation}
J =  \frac{1}{ L^3 \left< t_\mathrm{w} \right>}, \label{eq_nucleation_rate}
\end{equation}
with the linear size of the system $L$.
It is widely assumed that a bubble-nucleation process is a Poisson process, then 
the cumulative distribution function (CDF) associated with nucleation events has the exponential distribution
\begin{equation}
F(t) \equiv P(t_\mathrm{w} <t) = 1 - \exp(-t/\tau).  \label{eq_CDF}
\end{equation}
The exponential distribution is specified only by the parameter $\tau$,
which denotes the time scale of this stochastic process.
From Eq.~(\ref{eq_CDF}), $\left< t_\mathrm{w} \right>$ equals $\tau$.

On the other hand, CNT predicts the nucleation rate from the work carried out to form the critical bubble as
\begin{equation}
J = D Z c(n^*, v^*), \label{eq_nucleation_rate_cnt}
\end{equation}
where $D$ is the kinetic prefactor, $Z$ is the Zeldovich factor, which describes the nonequilibrium effect, and
$c(n,v)$ is the number density of bubbles with volume $v$ containing $n$ particles~\cite{Blander1975}.
In a superheated liquid, the reversible work $W(n,v)$ carried out to form a bubble with volume $v$ containing $n$ particles
has a saddle point at $(n^*, v^*)$, which corresponds to the critical bubble.
The nucleation rate $J$ given in Eq.~(\ref{eq_nucleation_rate_cnt}) is determined only by
the temperature and density of the superheated liquid.
Equations (\ref{eq_nucleation_rate}) and (\ref{eq_nucleation_rate_cnt}) lead to the finding that
$\left< t_\mathrm{w} \right>$ should be inversely proportional to the volume of the system
if the system does not have any finite-size effect.

\begin{figure}[bt]
\begin{center}
\includegraphics[width=\linewidth]{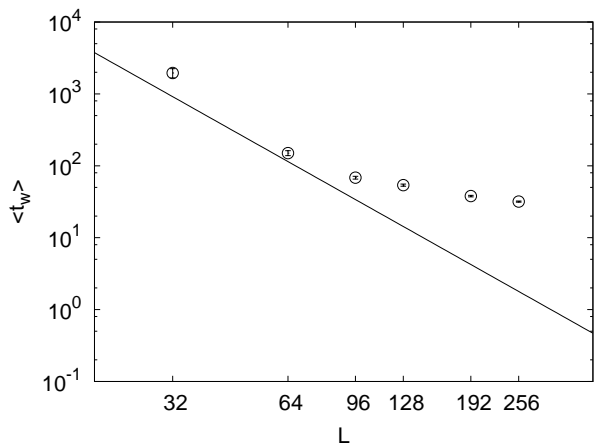}
\end{center}
\caption{
System-size dependence of the expectation value of the waiting time $\left< t_\mathrm{w} \right>$.
Decimal logarithms are taken for both axes.
The error bars are smaller than the symbols.
The solid line, which denotes $L^{-3}$, is a guide for the eyes.
}
\label{fig_nucleation_rate_a}
\end{figure}

\section{Method}

To estimate the waiting time in bubble nucleation,
we perform MD simulations.
We use the truncated Lennard-Jones potential of the form
\begin{equation}
V(r) = 
\displaystyle 4 \varepsilon \left[
\left( \frac{\sigma}{r} \right)^{12} -
\left( \frac{\sigma}{r} \right)^{6} +
c_2 \left( \frac{r}{\sigma} \right)^2+ c_0 \right],
\end{equation}
with the well depth $\varepsilon$ and atomic diameter $\sigma$~\cite{Spotswood1973}.
The coefficients $c_2$ and $c_0$ are determined so that
$V(r_c) = V'(r_c) = 0$ with the cut-off length $r_c$ , i.e., the values of potential and force 
become continuously zero at the truncation point.
In the following, we use the physical quantities reduced by $\sigma$, $\varepsilon$, and $k_\mathrm{B}$,
i.e., the length scale is measured using the unit of $\sigma$, and so forth.
We set the cutoff length as $r_c = 3.0$.
The system is a cube with linear size $L$ and is periodic in all directions.
The number of particles is chosen so that the initial density of the system $\rho \equiv N/L^3$ is $0.7$.
We first maintain the system in the pure-liquid phase using a thermostat,
then we expand the system.
The expansion is performed by changing the radius of the particles from $\sigma$ to $\sigma' = \alpha \sigma$, where $\alpha$ is 
a rescaling factor.
This procedure is equivalent to the the uniform and adiabatic expansion,
which is $\textbf{q}_i \rightarrow \textbf{q}_i/\alpha$ and $L \rightarrow L/\alpha$ where
$\textbf{q}_i$ is the position of the particle $i$.
Note that all physical quantities should be rescaled after expansion, since
we measure them in the unit of the radius $\sigma$, for example, $T' =T/\alpha^2$, $\rho' = \rho \alpha^3$, and so forth.
We chose the rescaling factor $\alpha$ to be $0.98$ for all runs, and therefore, the change in the density upon the expansion
 is from $\rho=0.7$ to $0.659$ \cite{temperaturedrop}.
After the expansion, we turn off the thermostat and continue the microcanonical simulation.
The system is thermalized at the temperature $T=0.9$ using the Nos\'e--Hoover method~\cite{NoseHoover}.
The integration scheme for the isothermal time evolution is the second-order Reversible System Propagator Algorithm (RESPA).~\cite{Tuckerman1992},
and the leapfrog algorithm is used for the microcanonical simulation with the time step $\Delta t =0.005$.
The typical time evolutions of temperature and pressure are shown in Fig.~\ref{fig_temperature}.
One can see that both temperature and pressure suddenly drop when the systems are expanded, and 
they relax to values for metastable states, which are superheated liquids with negative pressure.
The time evolutions of temperature and pressure are less affected by the size of the system.
All systems become superheated liquids at temperature $T=0.868(2)$.
We obtain the phase diagram of this system from the preliminary simulations.
We also determine the spinodal line between the liquid and liquid-gas coexistent phases
by the method described in Ref.~\cite{Kuksin}.
The obtained phase diagram is shown in Fig.~\ref{fig_phasediagram}.
As shown in the figure, the system with $\rho=0.659$ and $T=0.868$ is in the liquid-gas coexistent region.
The densities at the binodal and spinodal points for $T=0.868$ are $0.695(1)$ and $0.640(1)$, respectively.
In order to identify bubbles, we divide the system into small subcells with length $2.0$
and observe the local density for each subcell. From the preliminary simulations,
the densities of gas and liquid coexisting in this system at $T=0.9$ are estimated to be $0.04(1)$ and 
$0.67(2)$, respectively. Therefore, we define a subcell to be in the {\it gas state} when its density is less than $0.2$.
We have confirmed that the results do not change for other values of the threshold such as $0.3$.
We define that the neighboring {\it gas state} cells are in the same cluster and
identify the bubble using the site-percolation criterion in the simple cubic lattice.
The time evolution of a bubble identified by the above method is shown in Fig.~\ref{fig_nucleation}.

\begin{figure}[tb]
\begin{center}
\includegraphics[width=\linewidth]{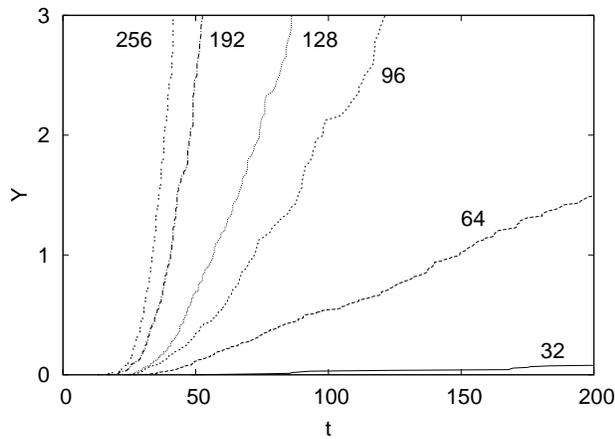}
\end{center}
\caption{
Cumulative distribution functions of nucleation events.
The values of $Y(t) = -\ln (1- F)$ are shown as functions of time.
The labels near lines denote system sizes.
}
\label{fig_logf}
\end{figure}

\section{Results}

We first estimate the critical size of a bubble.
After expansion, the volume of the largest bubble fluctuates for a certain period of time,
and then the monotonic development of a bubble with a different waiting time is observed.
If a bubble exceeds some critical size, then it starts to grow explosively.
Conversely, the sizes of bubbles before the explosive growth should be smaller than the critical size.
The maximum volume of the bubbles in the region of fluctuation is estimated to be about $v=180$.
We therefore define the waiting time $t_\mathrm{w}$ as the interval between the time of expanding operation ($t=50$)
and the time when the volume of the largest bubble reaches $v=200$.
We study several system sizes from $L=32$ to $256$.
The sizes of the studied systems and the number of particles are listed in Table~\ref{tbl_values}.
We observe 256 independent samples of the waiting time for each system size
and take their simple arithmetic mean.
Computations are mainly performed on HITACHI SR16000/L1 (32 ways on 1 node).
For the largest systems with 11744051 particles, 
40850 steps including those for thermalization are calculated in an average of 12274,
which give the calculation speed of 39.1 million updates per second.
The system-size dependence of the waiting time is shown in Fig.~\ref{fig_nucleation_rate_a}.
It is apparent from the figure that the waiting time is not proportional to $L^{-3}$,
and therefore the nucleation rate exhibits a strong finite-size effect.

\begin{figure}[tb]
\begin{center}
\includegraphics[width=\linewidth]{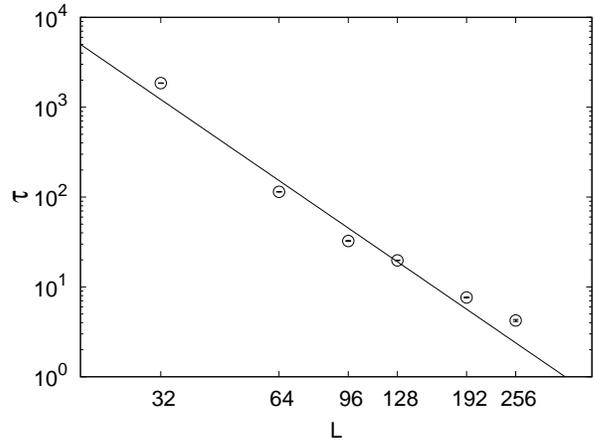}
\end{center}
\caption{
System-size dependence of the characteristic time scale $\tau$ of the Poisson process.
Decimal logarithms are taken for both axes.
The error bars are smaller than the symbols.
The solid line, which denotes $L^{-3}$, is a guide for the eyes.
}
\label{fig_nucleation_rate_s}
\end{figure}

In order to clarify the reason for the finite-size effect, we observe the 
probability distribution of the nucleation events.
First, we determine whether the nucleation process is a Poisson process.
Equation~(\ref{eq_CDF}) leads to
\begin{equation}
t/\tau = -\ln \left[1 - F(t) \right],
\end{equation}
which means that the function $Y(t) = -\ln \left[1 - F(t) \right]$ becomes linear 
and its slope gives the parameter of the distribution provided the nucleation is a Poisson process.
The values of $Y(t)$ are shown in Fig.~\ref{fig_logf}.
Whereas the lines are straight for small systems, those of larger systems bend at low values of $t$.
Additionally, the $x$-intercepts are not located at the origin, which suggests that the 
distribution of the waiting time has the form
\begin{equation}
F(t) = 
\left \{
\begin{array}{lc}
0 & \quad  t \leq t_0, \\
1 - \exp\left[-(t-t_0)/\tau\right] & \quad  t \geq t_0 ,
\end{array}
\right. \label{eq_CDF_real}
\end{equation}
with the additional relaxation time $t_0$.
Therefore, we perform a fitting assuming the form given by Eq.~(\ref{eq_CDF_real}).
For systems with $L \geq 96$, we only apply a fit to the region where the line appears to be straight in Fig.~\ref{fig_logf}.
The fitting results are summarized in Table \ref{tbl_values}.
It is shown that the additional relaxation time $t_0$ is almost independent of the system size.
The additional relaxation time was also reported in droplet-nucleation~\cite{Chkonia2009,Wedekind2009}.
While it was considered to be the time that the system needs to produce a nucleated cluster in droplet-nucleation,
it can be a result due to the expansion, since the relaxation time does not exhibits the finite-size effect
such as the temperature of the liquid under expansion as shown in Fig.~\ref{fig_temperature}.
In order to estimate the relaxation time due to the expansion, we observed the 
autocorrelation function of the temperature after the expansion, that is, the autocorrelation function of the 
superheated liquid.
We assumed that the autocorrelation function's form has a simple exponential form, $C(t) \sim \exp(-t/\tau_\mathrm{a})$,
with the characteristic time scale $\tau_\mathrm{a}$, and we found that $\tau_\mathrm{a} = 2.00(2)$,
which is too short to explain the value of additional relaxation time $t_0 \sim 30$.
We also check the definition of the waiting time $t_\mathrm{w}$.
The value of the waiting time depends on the volume of the critical bubble
where we choose $v = 200$ for the nucleation threshold in this study.
However, the value of $\tau$ should be independent of this definition
since it is a parameter which depends only on the state of the superheated liquid.
In order to confirm the above, 
we observe the waiting time with different values of the threshold $v = 180$ and $220$ for $L=64$.
Then we find that the additional waiting time $t_0$ becomes longer for larger values of the threshold
as $t_0 = 35.0, 36.6, $ and $37.4$ for $v= 180, 200,$ and $220$
while the timescale parameter $\tau $ is almost independent of the value of the threshold.
Therefore, we conclude that the additional relaxation time in bubble-nucleation is also 
a time to produce a nucleated cluster as in droplet-nucleation.

\begin{figure}[tb]
\begin{center}
\includegraphics[width=\linewidth]{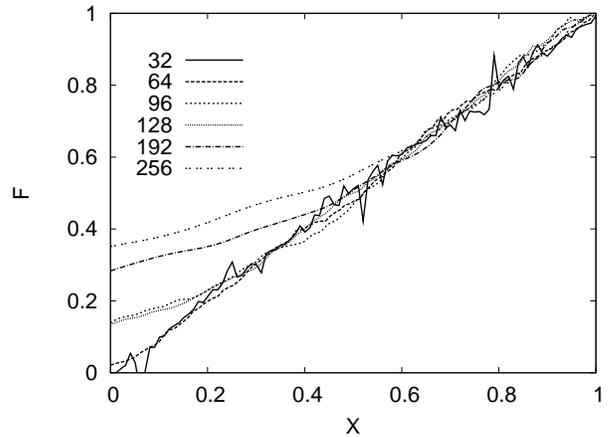}
\end{center}
\caption{
Cumulative distribution function $F(t)$ plotted as a function of
$X = 1 - \exp\left[ -(t-t_0)/\tau \right]$ for system sizes with the values listed in 
Table.~\ref{tbl_values}. If a nucleation process is perfectly described by Eq.~(\ref{eq_CDF_real}),
then this plot becomes a line connecting the origin to $(1,1)$.
}
\label{fig_scale}
\end{figure}

The system-size dependence of the parameter of the exponential distribution $\tau$ is shown in Fig.~\ref{fig_nucleation_rate_s},
which shows that the parameter is almost proportional to $L^{-3}$, which implies that the
nucleation rate is almost independent of the system size.
Assuming the relation $J = 1/(\tau L^3)$, as suggested by Fig.~\ref{fig_nucleation_rate_s}, we estimate the nucleation rate
to be $J = 1.66(5) \times 10^{-8}$, which is much larger than the value of $J \sim 2.63 \times 10^{-11}$ predicted by CNT, as well known~\cite{Zeng}.
The nucleation rate still exhibits the finite-size effect for larger systems,
for example, the value of $\tau$ when $L=256$ is only five times shorter than that when $L=128$,
which should be eight times shorter without the finite-size effect.
In order to illustrate how close the bubble nucleation is to a Poisson process,
we plot the CDF as a function of the value $X = 1 - \exp\left[-(t-t_0)/\tau\right]$
in Fig.~\ref{fig_scale}. If the nucleation process is a stochastic process 
with the distribution given by Eq.~(\ref{eq_CDF_real}), then the 
plot will become a straight line connecting the origin to $(1,1)$.
The figure shows that the CDFs of the larger systems deviate from
the Poisson process for small values of $X$.
This implies that the relaxation process caused by the expansion may affect bubble nuclei,
but it is difficult to separate the time scale of bubble nuclei from that due to the
expansion when two time scales are similar to each other.

\begin{table}[tb]
\begin{center}
\begin{tabular}{ccccccc}
\hline
$L$    & 32      & 64       & 96      & 128     & 192     & 256     \\
$N$    & 22937   & 183500   & 619315  & 1468006 & 4954521 &11744051 \\
$\left< t_\mathrm{w} \right> $ & $2.0(3)\times 10^3$ & 150(12) & 68(3) & 54(2) & 37.9(9) & 31.8(6) \\
$\tau$ & 1854(1) & 114.3(5) & 32.5(3) & 19.7(1) & 7.63(8) & 4.2(1) \\
$t_0$  & 27(4)   & 36.6(2)  & 39.7(2) & 36.3(2) & 32.4(1) & 29.6(1) \\ \hline
\end{tabular}
\end{center}
\caption{
Summary of the physical quantities for systems with length $L$ and number of particles $N$.
The arithmetic mean of the waiting time is denoted by $\left< t_\mathrm{w} \right>$.
The characteristic time of the exponential distribution $\tau$ and the 
additional relaxation time $t_0$ are determined by  the CDFs of $t_\mathrm{w}$.
}
\label{tbl_values}
\end{table}

\section{Summary and Discussion}

In the present study, we investigated the bubble-nucleation process of a Lennard-Jones liquid by MD simulations. 
We found that the nucleation rate, defined by the arithmetic mean of the waiting time,
exhibits an apparent finite-size dependence that can be removed by analysis of the CDF associated with the nucleation events.
The obtained CDF is found to be an exponential distribution with an extra delay time.
Therefore, it is necessary to study a system in which the waiting time is sufficiently longer than the additional relaxation time
to study bubble nuclei in a superheated liquid.
The nucleation rate correctly determined by CDF
has a smaller finite-size effect. This suggests that the discrepancy between the prediction of CNT and the experimental data
does not originate from the result of interbubble interactions though the pressure of the ambient liquid,
but from the inaccurate estimation of the
reversible work carried out to form a critical bubble.
To verify this conjecture, the direct measurement of the reversible work is required, 
which can be achieved by observing the bubble distribution in a superheated liquid.
This is one of important issues.
We investigated the origin of the additional relaxation time $t_0$.
Similar to the droplet-nucleation case, $t_0$ is the time required to make the critical bubble
from the candidates which are the fluctuating  embryos in the superheated liquid.
We also studied some different expansion ratios and found that the
additional relaxation time hardly depends on the expansion ratio
which implies that $t_0$ depends only on the density and the temperature of the meta-stable liquid after the expansion.
Figure \ref{fig_scale} shows that there are bubble nuclei in the short-time region $t < t_0$, which is apparent for larger systems,
while we assumed that the probability of nucleation is zero in Eq.~(\ref{eq_CDF_real}).
This suggests that the inhomogeneity of density caused by the expansion of the system may enhance bubble nuclei,
but further studies are required to clarify the effect of expansion on the formation of bubbles.

\section*{Acknowledgements}
The authors would like to thank S. Takagi and T. Komatsu for fruitful discussions and S. Sasa for valuable comments.
This work was partially supported by Grants-in-Aid for Scientific Research (Contract No.\ 19740235)
and by KAUST GRP (KUK-I1-005-04). 
The computation was partly carried out using the facilities of the Supercomputer Center,
Institute for Solid State Physics, University of Tokyo, and the Research Institute for Information Technology, Kyushu University.


\begin{thebibliography}{99}
\bibitem{temperaturedrop} When the system is expanded by changing the radii of the particles,
the temperature decreases since the positive work is required for the expansion.
In this simulation, the temperature just determined by the velocities of the particles
decreases from $T=0.9$ to $0.834$ by the expansion.
But we have to measure the temperature in terms of the new scale $\alpha \sigma$,
then the value of the temperature after the expansion is determined to be $0.868$.
\end{thebibliography}
\end{document}